\documentclass[a4paper,11pt]{article}
\pdfoutput=1
\usepackage{jheppub} 
\usepackage{tabularx, booktabs}
\usepackage{slashed}
\newcommand{\beq}{\begin{equation}}
\newcommand{\eeq}{\end{equation}}
\newenvironment{Eqnarray}{\arraycolsep 0.14em\begin{eqnarray}}{\end{eqnarray}}
\def\beqa{\begin{Eqnarray}}
\def\eeqa{\end{Eqnarray}}

\title{ Chasing the muon EDM to constrain the SMEFT and UV models}
\author[a]{Kuldeep Deka,}
\author[a]{Marta Losada,}
\author[b]{Yosef Nir}
\affiliation[a]{New York University Abu Dhabi\\
PO Box 129188, Saadiyat Island, Abu Dhabi, United Arab Emirates}
\affiliation[b]{Department of Particle Physics and Astrophysics\\
Weizmann Institute of Science, Rehovot 7610001, Israel}

\emailAdd{kuldeep.deka@nyu.edu}
\emailAdd{marta.losada@nyu.edu}
\emailAdd{yosef.nir@weizmann.ac.il}

\abstract{
Following a proposal for an experiment with sensitivity to an electric dipole moment (EDM) of the muon $d_\mu$ of order $6\times10^{-23}\ e$ cm, three to four orders of magnitude below the current bound, but still seven orders of magnitude above the current bound on the EDM of the electron $d_e$, we explore the discovery potential of such an experiment. Within the dimension-six CP violating operators of the Standard Model effective field theory (SMEFT), we identify two dipole operators where $d_\mu$ has the strongest sensitivity, and four classes of four-fermion operators where it has the best sensitivity for regions of parameter space that are far from minimal flavor violation. We further consider three UV completions: vector-like leptons (VLLs), heavy vector boson with off-diagonal leptonic couplings, and two Higgs doublet model. For each, we identify the region in parameter space that will be uniquely explored by the proposed $d_\mu$ experiment. In case of VLLs, we also find measurements of $\Gamma (h \rightarrow \mu \mu)$ offer competitive sensitivity, highlighting the complementary role of collider observables. Generically, the potential reach is to ${\cal O}(10\ {\rm TeV})$ scale of new physics.}

\begin{document}
\begin{flushright}
\end{flushright}
\maketitle
\section{Introduction}

The electric dipole moments (EDMs) of the electron and the muon are CP violating observables that provide a sensitive probe of new CP violating sources at high energy scales. The current experimental bound on $d_e$ is given by \cite{Roussy:2022cmp}
\begin{equation}\label{eq:deexpbound}
    |d_e|<4.1\times10^{-30}\ e\ {\rm cm}\ (90\%\ {\rm C.L.}).
\end{equation}
The current experimental bound on $d_\mu$ is given by \cite{Muong-2:2008ebm}
\begin{equation}
     |d_\mu|<1.9\times10^{-19}\ e\ {\rm cm}\ (95\%\ {\rm C.L.}).
\end{equation}
There are also indirect ways to constrain $d_\mu$ \cite{Feng:2001sq,Feng:2002wf}. For example, the CP-odd effects in ThO molecule imply $|d_\mu|<2\times10^{-20}\ e$ cm  \cite{Ema:2021jds}. The Standard Model (SM) predictions for these EDMs are $d_e\sim10^{-39}\ e\ {\rm cm}$ and $d_\mu\sim10^{-38}\ e\ {\rm cm}$ \cite{Yamaguchi:2020eub}.

Recently, a new experiment for $d_\mu$ was proposed \cite{Adelmann:2025nev, Crivellin:2018qmi}, with an expected sensitivity of
\begin{eqnarray}\label{eq:dmuexpsensitivity}
     \sigma(d_\mu)&<&3\times10^{-21}\ e\ {\rm cm}\ ({\rm Phase\ I}),\nonumber\\
     \sigma(d_\mu)&<&6\times10^{-23}\ e\ {\rm cm}\ ({\rm Phase\ II}).
\end{eqnarray}
The following question arises: What type of physics beyond the Standard Model (BSM) can be probed by a $d_\mu$ experiment, with the sensitivity presented in Eq.~(\ref{eq:dmuexpsensitivity}), which is beyond the reach of the $d_e$ bound of Eq. (\ref{eq:deexpbound})? The aim of this paper is to provide concrete, qualitative and quantitative, answers to this question.

Most of our analysis will be conducted in the framework of the Standard Model effective field theory (SMEFT), where dimension-six operators that are products of the SM fields, contracted to singlets of the SM gauge group, represent new sources of CP violation. In UV completions where the flavor structure is similar to that of the SM, one often obtains $|d_\mu/d_e| \sim m_\mu/m_e\sim200$ (see, for example, Ref.~\cite{Abada:2024hpb}), so that $d_\mu$ is orders of magnitude below the experimental sensitivity. There are, however, full high-energy models with a flavor structure such that the effects on $d_\mu$ are particularly enhanced, allowing (in part of the parameter space) $d_\mu$ within the range of Eq.~(\ref{eq:dmuexpsensitivity}). 

It has already been exhaustively studied how $d_e$ can provide the strongest constraints on BSM models. In addition, various specific UV models have been studied in the context of BSM enhancement of $d_\mu$. Ref.~\cite{Hou:2021zqq} studies a two Higgs doublet model (2HDM) with sizable off-diagonal $\mu-\tau$ Yukawa couplings. Ref.~\cite{Nakai:2022vgp} studies a muon-specific 2HDM, where a $Z_4$ symmetry guarantees that the muon couples exclusively to one of the Higgs doublets, while the electron and the tau-lepton couple to the other Higgs doublet. In a supersymmetric seesaw model of neutrino masses with non-degenerate heavy neutrinos \cite{Ellis:2001yza}, the neutrino Yukawa interactions induce large corrections to the slepton soft masses. Ref.~\cite{Hiller:2010ib} presents a supersymmetric model with a hybrid gauge-gravity breaking, where a large mixing between the $\mu$ and $\tau$ sleptons leads to large $d_\mu$ values. Ref.~\cite{Hamaguchi:2022byw} studies a model which contains vector-like leptons that couple only to second-generation SM leptons. In the 2HDM extended by vector-like leptons of Ref.~\cite{Dermisek:2023tgq}, a strong correlation between ${\rm BR}(h\to\mu^+\mu^-)$ and $d_\mu$ arises, so that a deviation of ${\cal O}(0.01)$ of the former requires large values for the latter. The model of Ref.~\cite{Khaw:2022qxh} contains a dark matter fermion and new scalars that couple exclusively to the muon. Leptoquark models that address $B$-anomalies can enhance $d_\mu$ significantly \cite{Altmannshofer:2020ywf}. We note that most of these models were motivated by the claimed discrepancy between the SM prediction for the anomalous magnetic moment of the muon and its experimental range. 

Three specific examples are reviewed in the following, with a comparison of the constraints from $d_e$, $\Delta a_\mu$, and ${\rm BR}(\mu\to e\gamma)$ with the constraint from the projected $d_\mu$ sensitivity (\ref{eq:dmuexpsensitivity}). 
For the muon $g-2$ constraint, we use the latest experimental value~\cite{Muong-2:2025xyk} and the theory value corresponding to the average of the lattice results~\cite{Aliberti:2025beg}, resulting in $\Delta a _\mu \lesssim 1 \times 10^{-9}$~\cite{Aliberti:2025beg}. For the lepton flavor violating (LFV) decay, we use $\text{BR}(\mu\to e\gamma) \lesssim 1.5 \times 10^{-13}$ from the MEG II experiment~\cite{MEGII:2025gzr}. 

The leptonic dipole operators are given by
\begin{equation}
\mathcal{L}_{\text{EDM}} = -\frac{i}{2} d_\ell(E)\, \bar{\ell} \sigma^{\mu\nu} \gamma_5 \ell\, F_{\mu\nu},
\label{edmdef}
\end{equation}
where $\ell=e,\mu,\tau$ and $F_{\mu\nu}$ is the electromagnetic field strength tensor. The EDM can be extracted from the Wilson coefficient of the dipole operator evaluated at the mass of the corresponding lepton: $d_e = d_e(E = m_e)$ or $d_\mu = d_\mu(E = m_\mu)$. 
To connect a high-scale new physics (NP) model at $\Lambda$ to this low-energy observable, one must run the operator coefficients down to the IR scale via renormalization group (RG) evolution. This RG flow proceeds through multiple EFT regimes: from the NP scale $\Lambda$ to the electroweak (EW) scale using the SMEFT, and from the EW scale to the lepton mass scale using a low-energy EFT (LEFT) containing only photons, gluons, and light fermions. 

The structure of this paper is as follows. In Section~\ref{sec:smeft} we list the dimension-six CP violating SMEFT operators that are relevant to our discussion and review their contributions to the leptonic EDMs. In Section~\ref{sec:results} we calculate and compare the bounds from $d_e$ and $d_\mu$ on the Wilson coefficients of the various operators. The bounds on the SMEFT operators are applied to specific UV completions in Section~\ref{sec:UV_dipoles}. We conclude in Section~\ref{sec:con}. 

\section{The SMEFT Contributions to the EDMs}
\label{sec:smeft}
The SMEFT operators relevant to leptonic EDMs have been studied in Refs.~\cite{Panico:2018hal,Aebischer:2021uvt,Kley:2021yhn}. We follow Ref.~\cite{Panico:2018hal} which listed the dimension-6 SMEFT operators contributing to the lepton EDMs at tree, one-loop and two-loop levels. We present their list in Table~\ref{operatortable}.
\begin{table}[h]
\centering
\begin{tabular}{|l|}
\multicolumn{1}{c}{\textbf{Tree-level}} \\
\hline
$\mathcal{O}_{\ell W} = (\overline{L_{L}} \sigma^a \sigma^{\mu\nu} \ell_R)\, H W^a_{\mu\nu}$ \\
$\mathcal{O}_{\ell B} = (\overline{L_{L}} \sigma^{\mu\nu} \ell_R)\, H B_{\mu\nu}$ \\
\hline
\multicolumn{1}{c}{\textbf{1-loop level}} \\
\hline
$\mathcal{O}_{luqe} = (\overline{L_{L}} u_R)(\overline{Q_L} \ell_{R})$ \\
$\mathcal{O}_{W\widetilde{W}} = |H|^2\, W^{a\, \mu\nu} \widetilde{W}^a_{\mu\nu}$ \\
$\mathcal{O}_{B\widetilde{B}} = |H|^2\, B^{\mu\nu} \widetilde{B}_{\mu\nu}$ \\
$\mathcal{O}_{W\widetilde{B}} = (H^\dagger \sigma^a H)\, W^{a\, \mu\nu} \widetilde{B}_{\mu\nu}$ \\
$\mathcal{O}_{\widetilde{W}} = \varepsilon_{abc} \widetilde{W}^a{}^\nu{}_\mu W^{b\rho}{}_\nu W^{c\mu}{}_\rho$ \\
\hline
\end{tabular}
\hspace{1cm}
\begin{tabular}{|l|}
\multicolumn{1}{c}{\textbf{2-loop level}} \\
\hline
$\mathcal{O}_{lequ}^{(1)} = (\overline{L_{L}} \ell_{R})(\overline{Q_L} u_R)$ \\
$\mathcal{O}_{e'W} = (\overline{L'_L} \sigma^a \sigma^{\mu\nu} \ell'_R)\, H W^a_{\mu\nu}$ \\
$\mathcal{O}_{e'B} = (\overline{L'_L} \sigma^{\mu\nu} \ell'_R)\, H B_{\mu\nu}$ \\
$\mathcal{O}_{uW} = (\overline{Q_L} \sigma^a \sigma^{\mu\nu} u_R)\, \widetilde{H} W^a_{\mu\nu}$ \\
$\mathcal{O}_{uB} = (\overline{Q_L} \sigma^{\mu\nu} u_R)\, \widetilde{H} B_{\mu\nu}$ \\
$\mathcal{O}_{dW} = (\overline{Q_L} \sigma^a \sigma^{\mu\nu} d_R)\, H W^a_{\mu\nu}$ \\
$\mathcal{O}_{dB} = (\overline{Q_L} \sigma^{\mu\nu} d_R)\, H B_{\mu\nu}$ \\
$\mathcal{O}_{le\bar{d}\bar{q}} = (\overline{L_L} \ell_R)(\bar{d}_R Q_L)$ \\
$\mathcal{O}_{le\bar{e}'\bar{l}'} = (\overline{L_L} \ell_R)(\overline{\ell'_R} L'_L)$ \\
$\mathcal{O}_{y_e} = |H|^2 \overline{L_L} \ell_R H$ \\
\hline
\end{tabular}
\caption{\it SMEFT operators contributing to the lepton EDM $d_\ell$ at different loop levels. $L_L(1,2)_{-1/2}$ and $\ell_R(1,1)_{-1}$ correspond to the lepton of interest ($\mu$ or $e$), while $L_L'$ and $\ell_R'$ to the other leptons. $Q_L(3,2)_{-1/6}$, $u_R(3,1)_{+2/3}$ and $d_R(3,1)_{-1/3}$ stand for the quark fields of any generation. }
\label{operatortable}
\end{table}

\subsection{Tree-level}
The tree-level lepton-EDM $d_\ell$ ($\ell=e$ or $\mu$) is generated directly by the dipole operators:
\begin{equation}
d_\ell = \frac{\sqrt{2}v}{\Lambda^2}\left[ s_{\theta_W} C_{\ell W}^I(E) - c_{\theta_W} C_{\ell B}^I(E) \right],
\end{equation}
where $s_{\theta_W}$ and $c_{\theta_W}$ are the sine and cosine of the Weinberg angle, respectively. 

\subsection{One-loop}
At one-loop, $d_\ell$ receives logarithmically enhanced contributions from CP-odd bosonic and four-fermion operators:
\begin{align}
\frac{d_\ell}{e} \simeq - \frac{1}{16\pi^2} \frac{\sqrt{2}v}{\Lambda^2} \Bigg[ y_u\, C_{luqe}^I 
- y_\ell \Bigg( 2 C_{B\widetilde{B}} c_{\theta_W}^2 + \frac{1 - 4 s_{\theta_W}^2}{s_{\theta_W} c_{\theta_W}} \left( s_{\theta_W} c_{\theta_W}(C_{W\widetilde{W}} - C_{B\widetilde{B}}) \right. \nonumber\\ 
\left. - \frac{1}{2}(c_{\theta_W}^2 - s_{\theta_W}^2) C_{W\widetilde{B}} \right) - \frac{C_{W\widetilde{B}}}{2 t_{\theta_W}} \Bigg) \Bigg] \ln\frac{\Lambda^2}{m_h^2}.
\label{deluqe}
\end{align}

$\mathcal{O}_{\widetilde{W}}$ can contribute to $\mathcal{O}_{\ell W}$ at one loop, but the contribution is finite:
\begin{equation}
    \frac{d_\ell}{e} \simeq \frac{3\,g}{32 \pi^2} \frac{y_\ell\, v}{\Lambda^2} C_{\widetilde{W}}
\end{equation}

\subsection{Two-loop: Double-log}
Double-logarithmic contributions at two loops arise from Barr--Zee-type diagrams and RG mixing:
\begin{align}
\frac{d_\ell}{e} &\simeq \frac{1}{(16\pi^2)^2} \frac{\sqrt{2}v}{\Lambda^2} \frac{1}{8} \text{Im} \Bigg[
g y_\ell \left( y_{e'}(11 + 9 t_{\theta_W}^2) C_{e'W} + 15 y_u(3 + t_{\theta_W}^2) C_{uW} + 3 y_d(3 + t_{\theta_W}^2) C_{dW} \right) \nonumber \\
& - 3 \frac{g}{t_{\theta_W}} y_\ell \left( y_{e'}(1 + 7 t_{\theta_W}^2) C_{e'B} - 3 y_u(1 + 7 t_{\theta_W}^2) C_{uB} + y_d(3 + 5 t_{\theta_W}^2) C_{dB} \right) \nonumber \\
& + 2 g^2 y_u (3 + 5 t_{\theta_W}^2) C_{lequ}^{(1)} + y_\ell g^3 (13 + 3 t_{\theta_W}^2) C_{\widetilde{W}} 
\Bigg] \ln^2 \frac{\Lambda^2}{m_h^2}.
\end{align}

\subsection{Two-loop: Single-log}
Subleading two-loop contributions involving Yukawa insertions and four-fermion operators yield the following:
\begin{align}
\frac{d_\ell}{e} \simeq -\frac{g^2}{(16\pi^2)^2} \frac{\sqrt{2}v}{\Lambda^2} \text{Im} \left[
\frac{3}{8} t_{\theta_W}^2 C_{y_\ell}
+ y_d \frac{1}{8} t_{\theta_W}^2 C_{le\bar{d}\bar{q}}
+ y_{e'} \frac{1}{8} (2 + 9 t_{\theta_W}^2) C_{le\bar{e}'\bar{l}'}
\right] \ln\frac{\Lambda^2}{m_h^2}.
\end{align}

\subsection{Finite Barr--Zee Contributions}
Finite Barr-Zee contributions arise from ${\cal O}_{y_f}$ operators, with $f=e,t,q$(light quarks) and $e^\prime$(heavy leptons).
\paragraph{From $C_{y_e}$:}
\begin{equation}
\frac{d_\ell}{e} \simeq - \frac{16}{3\sqrt{2}} \frac{e^2}{(16\pi^2)^2} v \left(2 + \ln\frac{m_t^2}{m_h^2}\right) \frac{C_{y_\ell}^I}{\Lambda^2}.
\end{equation}

\paragraph{From $C_{y_t}$:}
\begin{equation}
\frac{d_\ell}{e} \simeq - \frac{e^2}{(16\pi^2)^2} 2 \sqrt{2} N_c Q_t^2 \frac{m_\ell}{m_t} v \left(2 + \ln\frac{m_t^2}{m_h^2}\right) \frac{C_{y_t}^I}{\Lambda^2}.
\end{equation}

\paragraph{From light quarks $q$:}
\begin{equation}\label{eq:lightq}
\frac{d_\ell}{e} \simeq - \frac{e^2}{(16\pi^2)^2} 2 \sqrt{2} N_c Q_q^2 v \frac{m_\ell m_q}{m_h^2} \left( \ln^2\frac{m_q^2}{m_h^2} + \frac{\pi^2}{3} \right) \frac{C_{y_q}^I}{\Lambda^2}.
\end{equation}

\paragraph{From heavy leptons $e'$:}\footnote{Eqs.~(\ref{eq:lightq}) and (\ref{eq:heavyl}) correct for a factor of $\sqrt{2}$ missing in the corresponding equations in Ref.~\cite{Panico:2018hal}.}
\begin{equation}\label{eq:heavyl}
\frac{d_\ell}{e} \simeq - \frac{e^2}{(16\pi^2)^2} 2 \sqrt{2} Q_e^2 v \frac{m_\ell m_{e'}}{m_h^2} \left( \ln^2\frac{m_{e'}^2}{m_h^2} + \frac{\pi^2}{3} \right) \frac{C_{y_{e'}}^I}{\Lambda^2}.
\end{equation}

\section{Quantitative Results}
\label{sec:results}
We calculated the upper bounds from $d_e$ (\ref{eq:deexpbound}) and $d_\mu$ (\ref{eq:dmuexpsensitivity}) on the Wilson coefficients of the CP violating SMEFT operators of Table~\ref{operatortable}. The bounds are given in Table~\ref{operatortable_bounds}.

\begin{table}[t]
\centering
\small
\setlength{\tabcolsep}{4pt}
\begin{tabularx}{\textwidth}{@{}>{\centering\arraybackslash}l >{\centering\arraybackslash}X >{\centering\arraybackslash}l >{\centering\arraybackslash}X@{}}
\multicolumn{2}{c}{\textbf{Electron EDM Bound:} } &
\multicolumn{2}{c}{\textbf{Muon EDM Bound:} } \\
\multicolumn{2}{c}{($|d_e| < 4.1 \times 10^{-30}\, \mathrm{e\cdot cm}$)} &
\multicolumn{2}{c}{($|d_\mu| < 6 \times 10^{-23}\, \mathrm{e\cdot cm}$)} \\
\addlinespace[0.5ex]
\multicolumn{4}{c}{\textbf{Tree-level}} \\
\toprule
$C_{eW}$ & $2.1\times10^{-5}\, y_e g$ & $C_{\mu W}$ & $1.4\, y_\mu g$ \\
$C_{eB}$ & $2.1\times10^{-5}\, y_e g'$ & $C_{\mu B}$ & $1.4\, y_\mu g'$ \\
\midrule
\multicolumn{4}{c}{\textbf{One-loop}} \\
\midrule
$C_{luqe}$ & $3.7\times10^{-4}\, y_e y_t$ & $C_{luq\mu}$ & $25\, y_\mu y_t$ \\
$C_{W\widetilde{W}}$ & $1.8\times10^{-3}\, g^2$ & $C_{W\widetilde{W}}$ & $1.2\times10^2\, g^2$ \\
$C_{B\widetilde{B}}$ & $1.9\times10^{-3}\, g'^2$ & $C_{B\widetilde{B}}$ & $1.4\times10^2\, g'^2$ \\
$C_{W\widetilde{B}}$ & $8.9\times10^{-4}\, g g'$ & $C_{W\widetilde{B}}$ & $63\, g g'$ \\
$C_{\widetilde{W}}$ & $2.5\times10^{-2}\, g^3$ & $C_{\widetilde{W}}$ & $1.7\times10^3\, g^3$ \\
\midrule
\multicolumn{4}{c}{\textbf{Two-loop}} \\
\midrule
$C_{lequ}$ & $1.4\times10^{-2}\, y_e y_t$ & $C_{l\mu qu}$ & $9.6\times10^2\, y_\mu y_t$ \\
$C_{\tau W}$ & $93\, y_\tau g$ & $C_{\tau W}$ & $6.4\times10^6\, y_\tau g$ \\
$C_{\tau B}$ & $1.4\times10^2\, y_\tau g'$ & $C_{\tau B}$ & $9.6\times10^6\, y_\tau g'$ \\
$C_{\mu W}$ & $2.4\times10^4\, y_\mu g$ & $C_{e W}$ & $5.6\times10^7\, y_e g$ \\
$C_{\mu B}$ & $3.6\times10^4\, y_\mu g'$ & $C_{e B}$ & $1.1\times10^{14}\, y_e g'$ \\
$C_{tW}$ & $2.6\times10^{-3}\, y_t g$ & $C_{tW}$ & $1.8\times10^2\, y_t g$ \\
$C_{tB}$ & $4.8\times10^{-3}\, y_t g'$ & $C_{tB}$ & $3.3\times10^2\, y_t g'$ \\
$C_{bW}$ & $23\, y_b g$ & $C_{bW}$ & $1.5\times10^6\, y_b g$ \\
$C_{bB}$ & $17\, y_b g'$ & $C_{bB}$ & $1.1\times10^6\, y_b g'$ \\
$C_{le\bar d\bar q}$ & $3.7\, y_e y_t (y_t/y_b)$ & $C_{l\mu \bar d\bar q}$ & $2.7\times10^5\, y_\mu y_t (y_t/y_b)$ \\
$C_{le\bar e'\bar l'}$ & $0.23\, y_e y_t \left( \frac{y_t}{y_\tau} \text{ or } \frac{y_t}{y_\mu} \right)$ &
$C_{l\mu\bar\mu'\bar l'}$ & $1.7 \times 10^4\, y_\mu y_t \left( \frac{y_t}{y_\tau} \text{ or } \frac{y_t}{y_e} \right)$ \\
$C_{y_e}$ & $1.3\, y_e$ & $C_{y_\mu}$ & $8.9\times10^4\, y_\mu$ \\
\midrule
\multicolumn{4}{c}{\textbf{Two-loop finite}} \\
\midrule
$C_{y_e}$ & $0.78\, y_e$ & $C_{y_e}$ & $1.8\times10^{13}\, y_e$ \\
$C_{y_t}$ & $0.78\, y_t$ & $C_{y_t}$ & $5.3\times10^4\, y_t$ \\
$C_{y_b}$ & $1.5\times10^2\, y_b$ & $C_{y_b}$ & $1.0\times10^7\, y_b$ \\
$C_{y_\tau}$ & $1.8\times10^2\, y_\tau$ & $C_{y_\tau}$ & $1.2 \times 10^7\, y_\tau$ \\
$C_{y_\mu}$ & $1.8\times10^4\, y_\mu$ & $C_{y_\mu}$ & $5.3\times10^4\, y_\mu$ \\
\bottomrule
\end{tabularx}
\vspace{2mm}
\caption{
Upper limits on Wilson coefficients from electron and muon EDMs. Effective scale is $\Lambda = 10$ TeV. Couplings such as $y_f$, $g$, $g'$ indicate their dependence on the corresponding SM parameters.
}
\label{operatortable_bounds}
\end{table}

Comparing the constraints from the measurements of $d_e$ and $d_\mu$, we can divide the SMEFT operators to three categories:
\begin{enumerate}
    \item $d_\mu$ will give a stronger constraint:
    \begin{equation}
        C_{\mu W}(1.8\times10^4),\ \ C_{\mu B}(2.6\times10^4),
    \end{equation}
    where the number in parentheses is the ratio between the current $d_e$ bound and the projected $d_\mu$ bound.
\item $d_\mu$ and $d_e$ will constrain the SMEFT operators whose contributions depend on $y_\mu$ and $y_e$, respectively:
\begin{equation}
    C_{luq\mu}(6.8\times10^3),\ \ C_{l\mu qu}(6.7\times10^4),\ \ C_{l\mu\bar d\bar q}(7.2\times10^4),\ \ C_{l\mu\bar\mu'\bar l'}(7.4\times10^4),
\end{equation}
where the number in parentheses is how much weaker is the $d_\mu$ bound on an operator involving $\mu$ than the $d_e$ bound on the corresponding operator involving $e$. If this number is larger than $y_\mu/y_e\sim200$ then, in minimal lepton flavor violation (MLFV) models \cite{Cirigliano:2005ck,Cirigliano:2006su,Dery:2013aba}, the $d_e$ bound is more constraining.
\item $d_e$ will give a stronger constraint on all other operators.
\end{enumerate}

\section{UV Completions for Dipole Operators}
\label{sec:UV_dipoles}
The observables of interest --- the charged lepton anomalous magnetic moment $\Delta a_\ell$, electric dipole moment $d_\ell$, and radiative LFV decay rates ${\rm BR}(\ell_i\to\ell_j\gamma)$ --- are dominantly controlled by dipole operators in the low-energy effective theory (LEFT), that incorporate electroweak symmetry breaking (EWSB):
\beq
\mathcal{L}_{\rm eff}\supset \frac{e}{8\pi^2}\,C_{ji}\,(\bar\ell_j\sigma^{\mu\nu}P_R\ell_i)F_{\mu\nu}+\mathrm{h.c.},
\eeq
where $C_{ij}/(8\pi^2)$ is the effective dipole Wilson coefficient. 
These low energy operators arise from the SMEFT operators:
\begin{equation}
\mathcal O_{eB}^{ij} = (\overline{L}_i \sigma^{\mu\nu} e_j) H B_{\mu\nu},
\qquad
\mathcal O_{eW}^{ij} = (\overline{L}_i \sigma^{\mu\nu} e_j) \tau^I H W^I_{\mu\nu}.
\label{eq:SMEFT_dipoles}
\end{equation}

The diagonal components $C_{ii}$ determine $\Delta a_i$ and $d_i$, while the off-diagonal components $C_{ij}$ control $\ell_i \to \ell_j \gamma$:
\begin{align}
\Delta a_i &= -\frac{m_i}{2\pi^2}\,\text{Re}(C_{ii}),
\label{eq:univ_g2} \\[4pt]
d_i &= -\frac{e}{4\pi^2}\,\text{Im}(C_{ii}),
\label{eq:univ_edm} \\[4pt]
\text{BR}(\ell_j\to\ell_i\gamma) &= 
\frac{3\alpha_{\rm em}}{\pi G_F^2 m_{\ell_j}^2}\,\big(|C_{ji}|^2+|C_{ij}|^2\big).
\label{eq:univ_lfv}
\end{align}
In this section, we illustrate how these operators arise in three representative UV scenarios:  
vector-like leptons, heavy neutral gauge bosons, and heavy scalar doublets.

\subsection{Heavy fermions: Vector-like leptons (VLLs)}
\label{sec:VLL}
As an example for heavy fermions that can contribute to a leptonic EDM, consider a model with leptons in two different vector-like representations: $L'_{L,R}(1,2)_{-1/2}$ and $E^\prime_{L,R}(1,1)_{-1}$. The Lagrangian terms involving $L^\prime$  and $E^\prime$ are the following:
\begin{align}
\mathcal{L}_{\rm VLL} \supset{}& 
\overline{L'}(i\slashed{D}-M_L)L' + \overline{E^\prime}(i\slashed{D}-M_E)E^\prime
\nonumber\\
&\; - \Big[ 
\lambda_E^{\,i}\,\bar{\ell}_i H E^\prime_R 
+ \lambda_L^{\,i}\,\overline{L_L'} H e_{R}^i
+ \lambda\,\overline{L'_L} H E^\prime_R 
+ \bar\lambda\, H^\dagger \overline{E^\prime_L} L'_R 
+ \mathrm{h.c.}\Big],
\label{eq:VLL_lag}
\end{align}
where $i$ runs over SM flavors and $H=(0,(v+h))^T/\sqrt{2}$ with $v\simeq 246$ GeV.

Integrating out the $L^\prime$ and $E^\prime$ fields generates at tree level the ${\cal O}_{y_\mu}$ operator, with Wilson coefficient
\begin{equation}
\frac{C_{y_\mu}}{\Lambda^2}\;\simeq\;
-\,\frac{\lambda_E^\mu\,\bar\lambda\,\lambda_L^\mu}{M_E M_L}\,,
\label{eq:Cy_matching}
\end{equation}
and at one-loop the dipole operators $\mathcal O_{eW}$ and $\mathcal O_{eB}$, with Wilson coefficients
\begin{equation}
\frac{C_{eB}^{ji}}{\Lambda^2}\;\simeq\;
-\,\frac{e}{32 \pi^2  c_{\theta_W}}\;\frac{\lambda_E^j\,\bar\lambda\,\lambda_L^i}{M_E M_L}\,, \, \, \,
\frac{C_{eW}^{ji}}{\Lambda^2}\;\simeq\;
\,\frac{e}{32 \pi^2 s_{\theta_W}}\;\frac{\lambda_E^j\,\bar\lambda\,\lambda_L^i}{M_E M_L}\,.
\label{eq:Cy_matching}
\end{equation}

Since ${\cal O}_{y_\mu}$ contributes to the dipole operator ${\cal O}_{ij}$ at two loops, while ${\cal O}_{eB}$ and ${\cal O}_{eW}$ at tree level, the latter dominate, giving
\begin{equation}
C_{ji}^{\rm VLL} \;\simeq\; 
-\,\frac{v}{4\sqrt{2}}\;\frac{\lambda_L^i\,\lambda_E^j\,\bar\lambda}{M_L M_E}.
\label{eq:C_VLL}
\end{equation}

Using Eqs.~\eqref{eq:univ_g2}–\eqref{eq:univ_lfv}, we obtain the VLLs contributions to the anomalous magnetic moment, the electric dipole moment, and radiative LFV decay rate:
\begin{align}
\Delta a_i &\simeq 
+\frac{m_i v}{8\sqrt{2}\pi^2}\,
\text{Re}\!\left(\frac{\lambda_L^i\,\lambda_E^i\,\bar\lambda}{M_L M_E}\right),
\label{eq:VLL_g2}\\[4pt]
d_i &\simeq 
+\frac{e v}{16\sqrt{2}\pi^2}\,
\text{Im}\!\left(\frac{\lambda_L^i\,\lambda_E^i\,\bar\lambda}{M_L M_E}\right),
\label{eq:VLL_edm}
\end{align}
\begin{equation}
\text{BR}(\mu\to e\gamma)\;\simeq\;
\frac{3\alpha_{\rm em}\,v^2}{32\pi G_F^2 m_\mu^2}\,
\frac{|\bar\lambda|^2}{(M_L M_E)^2}
\Big(|\lambda_L^\mu\,\lambda_E^e|^2+|\lambda_L^e\,\lambda_E^\mu|^2\Big).
\label{eq:VLL_lfv}
\end{equation}

The complex muon Yukawa coupling, generated by the VLLs, contributes via a two-loop Barr-Zee diagram to the electron EDM~\cite{Panico:2018hal}:
\begin{equation}
\frac{d_e^{(\mu)}}{e} \simeq \frac{e^2}{64 \pi^4} \frac{v\, m_e m_\mu}{m_h^2} \frac{\text{Im}(\lambda_E^\mu \bar\lambda \lambda_L^\mu)}{M_L M_E} \left[ \ln^2 \left(\frac{m_\mu^2}{m_h^2}\right) + \frac{\pi^2}{3} \right]
\end{equation}

\begin{figure}[t!]
    \def\sepf{0.49}
    \centering
    \includegraphics[width=\sepf\columnwidth]{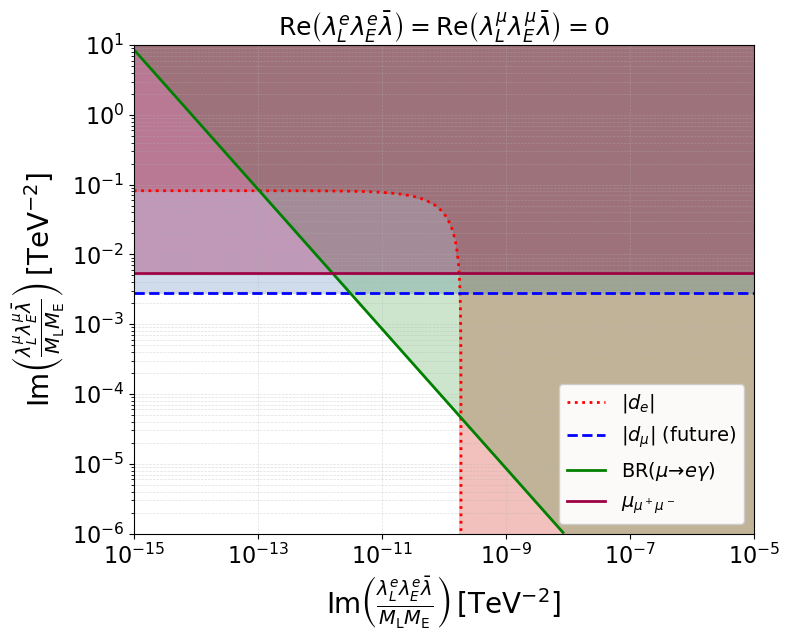}
     \includegraphics[width=\sepf\columnwidth]{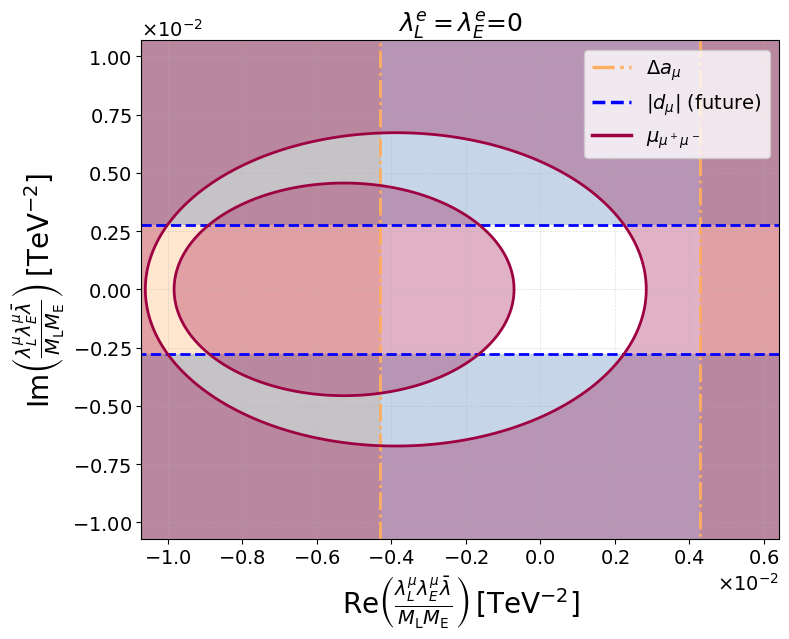}
    \caption{Constraints on the VLL model parameters $(\lambda_L\bar\lambda\lambda_E)/(M_LM_E)$ from the current ranges of $\Delta a_\mu$, $d_e$, ${\rm BR}(\mu\to e\gamma)$ and ${\rm BR}(h\to\mu^+\mu^-)$, and the projected reach of $d_\mu$. (Left) The $e-\mu$ plane. For $\text{BR}(\mu\to e\gamma)$, we assume $\lambda_L^e\lambda_E^e = \lambda_L^\mu\lambda_E^\mu$. (Right) The ${\rm Re}\frac{\lambda_L^\mu\bar\lambda\lambda_E^\mu}{M_LM_E}-{\rm Im}\frac{\lambda_L^\mu\bar\lambda\lambda_E^\mu}{M_LM_E}$ plane.}
    \label{vll_new}
\end{figure}

The constraints from $\Delta a_\mu$, $d_\mu$, $d_e$ and ${\rm BR}(\mu\to e\gamma)$ are illustrated in Fig.~\ref{vll_new}. In the left plot, we assume for the ${\rm BR}(\mu\to e\gamma)$ constraint that $\lambda_L^\mu\lambda_E^\mu=\lambda_L^e\lambda_E^e$, and for the $\Delta a_\ell$ and $d_\ell$ constraints that ${\rm Im}(\lambda_L^\ell\lambda_E^\ell)={\rm Re}(\lambda_L^\ell\lambda_E^\ell)$. In the right plot, we take $\lambda_{L,E}^e=0$. 

While $C_{y_\mu}$ affects $\Delta a_\mu$ and $d_\mu$ only at the two loop level, it modifies $\Gamma(h\to\mu\mu)$ at tree level, leading to strong constraints \cite{Fuchs:2019ore,Nir:2024oor}. The relevant observable is
\beq
\mu_{\mu^+\mu^-}\equiv\frac{\Gamma(h\to\mu^+\mu^-)}{[\Gamma(h\to\mu^+\mu^-)]^{\rm SM}},
\eeq
where $[\Gamma(h\to\mu^+\mu^-)]^{\rm SM}$ stands for the decay rate assuming the SM relation between the mass and the Yukawa coupling of the muon. 

The formalism for analyzing these constraints can be found in Ref.~\cite{Fuchs:2020uoc}. Defining
\beq
T_\mu\equiv-\frac{C_{y_\mu}v^2}{2y_\mu\Lambda^2}=\frac{\lambda_E^\mu\bar\lambda\lambda_L^\mu}{2y_\mu}\frac{v^2}{M_L M_E},
\eeq
we obtain:
\beq
\mu_{\mu^+\mu^-}=\frac{1+6{\rm Re}T_\mu+9|T_\mu|^2}{1+2{\rm Re}T_\mu+|T_\mu|^2}.
\eeq

The current experimental range is given by~\cite{CMS:2020xwi,ATLAS:2020fzp}:
\begin{equation}\label{eq:mumumu}
\mu_{\mu^+\mu^-}=1.21\pm0.35,
\end{equation}

The resulting constraints are shown in Fig.~\ref{vll_new}. We learn that, for vector-like leptons that couple dominantly to muons, the constraints from $\Delta a_\mu$, $\Gamma(h\to\mu^+\mu^-)$ and the projected $d_\mu$ are complimentary and competitive. For couplings of ${\cal O}(1)$, the sensitivity of all three measurements is to VLL masses up to ${\cal O}(20\ {\rm TeV})$.

\subsection{Heavy vector bosons: $U(1)_{L_2-L_3}$ gauge boson $Z'$}
\label{sec:Zprime}
As an example for a heavy vector boson that can contribute to a leptonic EDM, consider a model with a $U(1)_{L_2-L_3}$ gauge symmetry. Here, the $SU(2)$-doublet left-handed leptons $L_{L1}$, $L_{L2}$,and $L_{L3}$, and the $SU(2)$-singlet right-handed leptons $e_{R1}$, $e_{R2}$ and $e_{R3}$, are the interaction eigenstates with well-defined charges, $(0, +1,-1)$, under $U(1)_{L_2-L_3}$. It is important that they are close to, but not exactly, the mass eigenstates $e$, $\mu$ and $\tau$. In the space of the EM charge $-1$ leptons, $\ell_1$, $\ell_2$ and $\ell_3$, the gauge boson $Z^\prime$ couplings are given by
\beq
g_{Z^\prime}{\rm diag}(0,+1,-1).
\eeq
The $U(1)_{L_2-L_3}$ symmetry is spontaneously broken by the VEV of a SM-singlet scalar field $\Phi$ of charge $+1$ under the new symmetry. It provides the $Z^\prime$ vector boson with a mass-squared $M_{Z^\prime}^2$. The charged lepton mass matrix has two sources: The SM Yukawa couplings $y_{ij}$ of the Higgs doublet $H$ which are diagonal, and dimension-five terms of the form $(y^\prime_{ij}/\Lambda)\Phi H \overline{L_{Li}}e_{Rj}$,  $(ij)=(12),(13),(21),(31)$. ($\Lambda$ can be the mass scale of very heavy vector-like leptons.)  These two sources lead to the following charged lepton mass matrix:  
\beq
M_\ell=\frac{v}{\sqrt2}\begin{pmatrix}y_{11} & \epsilon y^\prime_{12} & \epsilon y^\prime_{13} \\\epsilon y^\prime_{21} & y_{22} & 0 \\ \epsilon y^\prime_{31} & 0 & y_{33} \end{pmatrix},
\eeq
where $\epsilon\equiv\langle\Phi\rangle/\Lambda\ll1$. Thus, the off-diagonal mass terms are in general small and the unitary matrices that transforms from the charged lepton interaction basis to the mass basis,
\beq
U^\ell_L M_\ell U^{\ell\dagger}_R={\rm diag}(m_e,m_\mu,m_\tau),
\eeq
are close to a unit matrix, but have small mixing angles, $s_{ij}^L$ and $s_{ij}^R$. Consequently, there are off-diagonal $Z^\prime$ couplings in the mass basis.

Integrating out the $Z^\prime$ field generates at one loop the dipole operators, with Wilson coefficients $C_{ij}/8\pi^2$:
\begin{equation}
    C_{ij} = -\frac{g_{Z'}^2}{M_{Z'}^2}\sum_k\Big[
    \big(s^R_{ik}s^{R*}_{jk}\,m_i 
       + s^L_{ik}s^{L*}_{jk}\,m_j\big)\,I_3(x_k)
    + s^L_{ik}s^{R*}_{jk}\,m_k\,I_4(x_k)
    \Big],
    \label{eq:C_Zprime}
\end{equation}
with $x_k=m_k^2/M_{Z'}^2$ and loop functions
\begin{align}
I_3(x) &= \frac{-8 + 38x - 39x^2 + 14x^3 -5x^4 + 18 x^2\ln x}{12(1-x)^4}\ \xrightarrow{x \to 0}\ -\frac{2}{3},
\label{eq:I3}\\
I_4(x) &= \frac{4 - 3x - x^3 + 6x\ln x}{2(1-x)^3}\ \xrightarrow{x \to 0}\ 2.
\label{eq:I4}
\end{align}

Using Eqs. (\ref{eq:univ_g2}) and (\ref{eq:univ_edm}), and taking the limit $x_i\ll1$ ($i=e,\mu,\tau$), we obtain the $Z^\prime$ contribution to the lepton anomalous magnetic moment and electric dipole moment:
\begin{align}
\Delta a_i &= 
+\frac{m_i g_{Z'}^2}{3\pi^2 M_{Z'}^2}\,
\sum_k\text{Re}\ \!\Big[-(|s^R_{ik}|^2+|s^L_{ik}|^2)\,m_i
+3s^L_{ik}s^{R*}_{ik}\,m_k\Big],
\label{eq:Zp_g2}\\[4pt]
d_i &= 
+\frac{e g_{Z'}^2}{2\pi^2 M_{Z'}^2}\,
\sum_k\text{Im}\ \!\Big( s^L_{ik}s^{R*}_{ik}\,m_k\Big).
\label{eq:Zp_edm}
\end{align}

Using Eq.~(\ref{eq:univ_lfv}), we obtain the $Z^\prime$ contribution to the $\mu\to e\gamma$ decay rate:
\beq
\text{BR}(\mu\to e\gamma) = 
\frac{3\alpha_{\rm em}}{\pi G_F^2 m_\mu^2}\frac{g_{Z^\prime}^4}{M_{Z^\prime}^4}\,\sum_k
\left[\left|-(2/3)
    \big(s^R_{ek}s^{R*}_{\mu k}\,m_e 
       + s^L_{ek}s^{L*}_{\mu k}\,m_\mu\big)
    + 2s^L_{ek}s^{R*}_{\mu k}\,m_k
    \right|^2+(e\leftrightarrow\mu)\right].
\label{eq:Zp_lfv}
\eeq

\begin{figure}[t!]
    \def\sepf{0.59}
    \centering
    \includegraphics[width=\sepf\columnwidth]{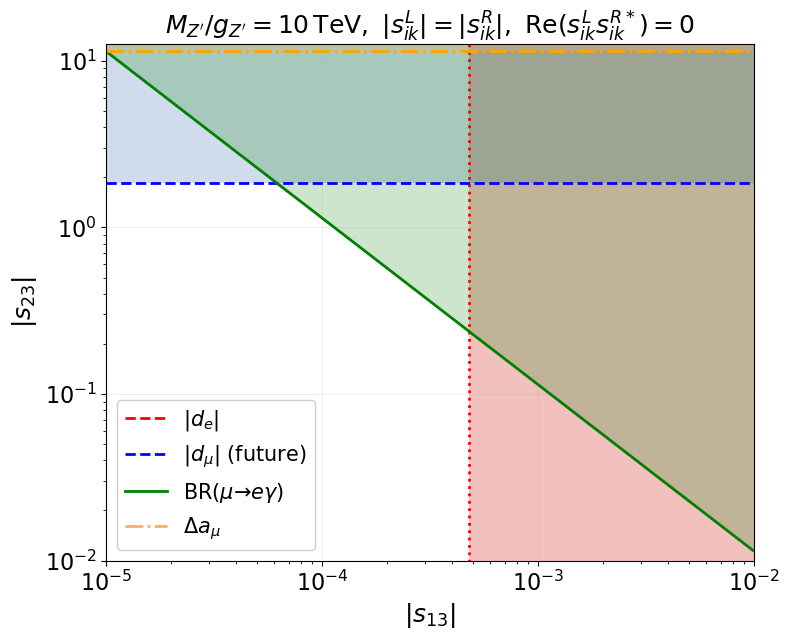}
    \caption{Exclusion regions from current bounds on $d_e$, BR($\mu\to e\gamma$) and $\Delta a_\mu$, and the region covered by the projected $d_\mu$, in the plane of lepton flavor changing couplings of a heavy vector boson $Z^\prime$. The $Z^\prime$ couplings to a pair of charged leptons $\overline{\ell}_i\ell_j$ of chirality $M$ is given by $g_{Z^\prime}s_{ij}^M$. The ratio $M_{Z^\prime}/g_{Z^\prime}$ is fixed at 10 TeV.} 
    \label{zp_new}
\end{figure}

In Fig.~\ref{zp_new}, we plot illustrative constraints in the $s_{13}-s_{23}$ plane, taking $(g_{Z^\prime}/M_{Z^\prime})=(1/10\ {\rm TeV})$, $|s^L_{ik}| = |s^R_{ik}|$ and ${\rm Re}(s^L_{ik}s^{R*}_{ik}) = 0$. We learn that $d_\mu$ will explore the regions of parameter space where $Z^\prime$ couples to $\mu\tau$ much more strongly than to $e\tau$, $|s_{13}|\ll|s_{23}|={\cal O}(1)$, and where the phase of the $Z^\prime\mu\tau$ coupling is ${\cal O}(1)$.

\subsection{Heavy Scalar: Extra Higgs Doublet}
\label{subsec:edm-lfv-g2-2hdm} 
As an example for a heavy scalar that can contribute to a leptonic EDM, consider a generic Two-Higgs-Doublet Model (2HDM). In such a model, there are two new sources of CP violation: Complex Yukawa couplings and mixing between the CP-even and CP-odd neutral scalars.

It is useful to write the Yukawa couplings in the basis $(\Phi_M,\Phi_A)$, where $\langle\Phi_M\rangle=v$ and $\langle\Phi_A\rangle=0$ (see, for example, \cite{Dery:2013aba}). We focus on the charged lepton Yukawa matrices:
\begin{equation}
-\mathcal{L}_Y^\ell= 
\overline{L_L} (Y_M^\ell \Phi_M + Y_A^\ell \Phi_A) E_R
+ \text{h.c.}.
\end{equation}
In the mass basis,
\beq
Y_M^\ell=(\sqrt{2}/v){\rm diag}(m_e,m_\mu,m_\tau),\ \ \ Y_A^\ell=\rho,
\eeq
where $\rho$ is a general complex matrix.

The angle between the $(\Phi_M,\Phi_A)$ basis and the mass basis for the neutral CP-even Higgs bosons, $(\Phi_H,\Phi_h)$, is commonly denoted by $\beta-\alpha$. The alignment (or decoupling) limit corresponds to
\begin{equation}
\sin(\beta-\alpha)=1. 
\end{equation}
In this limit, the Yukawa couplings of the light scalar $h$ are given by $Y_M$, reproducing the Standard Model Higgs couplings, while the Yukawa couplings of the heavy neutral scalars $H$ (CP-even) and $A$ (CP-odd) are given by
\begin{equation}
\mathcal{L}^\ell_{H,A} = -\frac{1}{\sqrt{2}}\, 
\overline{\ell_{Li}}\left(H - i\, A \gamma_5\right)\rho_{ij}\ell_{Rj}+ \text{h.c.}
\label{eq:LHA}
\end{equation}

In a general 2HDM, the scalar potential is complex, and the neutral scalar mass eigenstates are not CP eigenstates. We consider a small mixing between the CP-even and CP-odd scalars, and continue to call the neutral mass eigenstates $(h,H,A)$, even though they acquire small admixtures of opposite CP parity. We parameterize the CP violation by a dimensionless quantity $\varepsilon \ll 1$ that gives the CP-even component in the mass eigenstate $A$. The CP-odd component in $h$ is suppressed by both $\cos(\beta-\alpha)$ and $\varepsilon$, and we neglect it.  

The dominant one-loop contribution to the dipole operators comes from chirality flip of the internal lepton~\cite{Vives:2025clr,Altmannshofer:2020shb,Altmannshofer:2025nsl}:
\begin{equation}
C_{ij}^{(1)} \simeq 
m_k\,\rho_{ik} \rho_{kj}  \left(\frac{\cos^2 (\beta -\alpha)}{2 M_h^2 } \, I_2(x_{kh}) \,+\, \frac{\sin^2 (\beta-\alpha)}{2 M_H^2 } \,  I_2(x_{kH}) \,+\, \frac{1}{2 M_A^2}I_2(x_{kA}) \right) \,, 
\end{equation}
where $x_{ks}=m_k^2/M_s^2$ and
\beq
I_2(x) = \frac{3 - 4x + x^2 + 2\ln x}{8(1-x)^3}\ \xrightarrow{x \to 0}\ \frac{3}{8} + \frac{\ln x}{4}.  
\eeq

The dominant two-loop contribution to the dipole operators comes from Barr--Zee diagrams involving a top-quark loop~\cite{Vives:2025clr,Altmannshofer:2020shb,Altmannshofer:2025nsl}:
\begin{align}
\label{eq:2HBarrZeeCP}
C_{ij}^{t} 
&\simeq -\,\frac{3\,\alpha_{\rm em}\,Q_t^2}{2\sqrt{2}\pi v}
     \rho_{ij} \frac{\rho_{tt}}{m_t/v}\,\!
    \left\{\left[\, f(x_{tH}) -\, g(x_{tA}) \right]
    +\, i\,\varepsilon\, \left[\, f(x_{tH}) +\, g(x_{tA}) \right] \right\},
\end{align}
where 
\begin{align}
f(x) &= \frac{x}{2}\!\int_0^1 \!dz \, 
\frac{1 - 2z(1-z)}{z(1-z) - x}\,
\ln\!\frac{z(1-z)}{x}\ \xrightarrow{x \to 0}\ \frac{x}{2} \left( \ln x \right)^2 \nonumber\\
g(x) &= \frac{x}{2}\!\int_0^1 \!dz \, 
\frac{1}{z(1-z) - x}\,
\ln\!\frac{z(1-z)}{x}\ \xrightarrow{x \to 0}\ \frac{x}{2}  \left( \ln x \right)^2.
\end{align}

In the limit that the heavy scalars are much heavier than the electroweak scale, we have $M_S\equiv M_A\simeq M_H\gg v$, and the dipole operators are given by
\begin{align}
C_{ij}^{\text{tot}}
&\;\simeq\;
\underbrace{
\frac{m_k\,\rho_{ik}\rho_{kj}}{M_S^2}
\left(
  \frac{3}{8} + \frac{1}{4}\ln\frac{m_k^2}{M_S^2}
\right)
}_{\text{one-loop (internal flip)}}
-\;
\underbrace{
i\,\varepsilon\,\frac{3\,\alpha_{\rm em}\,Q_t^2}{2\sqrt{2}\,\pi}\,
\rho_{ij} \rho_{tt}\, 
   \frac{m_t}{M_S^2}
    \ln^2\frac{m_t^2}{M_S^2}
}_{\text{two-loop Barr--Zee (top loop)}},
\label{eq:CtotApprox}
\end{align}

\begin{figure}[t!]
    \def\sepf{0.49}
    \centering
    \includegraphics[width=\sepf\columnwidth]{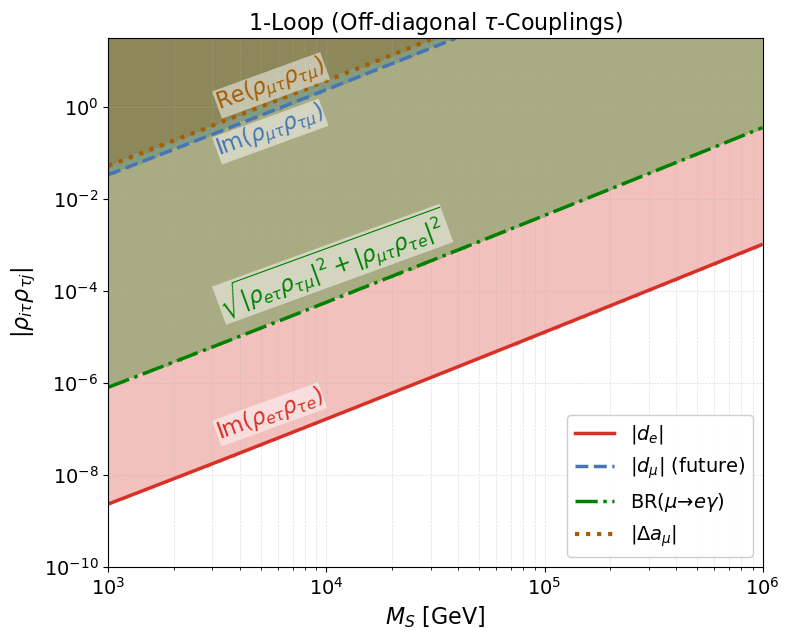}
    \includegraphics[width=\sepf\columnwidth]{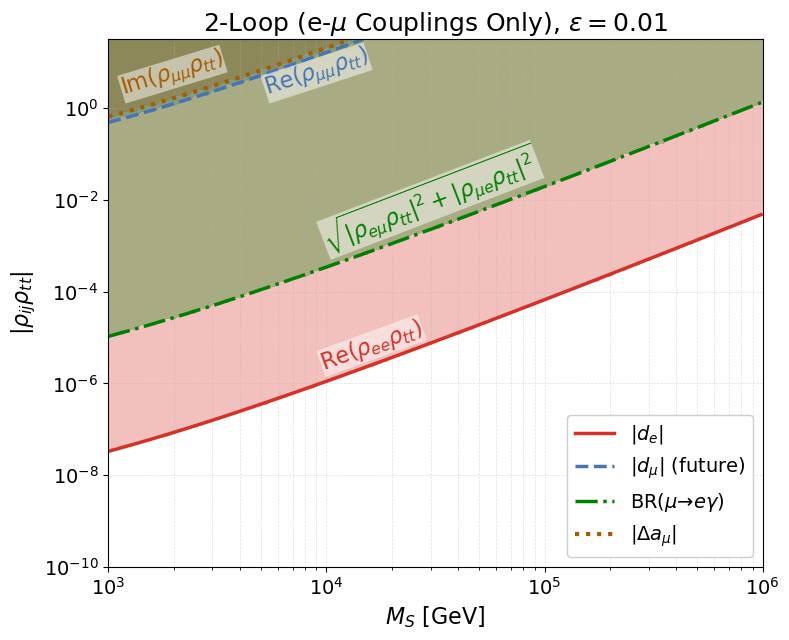}
    \caption{2HDM: Constraints from $d_e$, ${\rm BR}(\mu\to e\gamma)$, $\Delta a_\mu$ and (the projected) $d_\mu$ in the plane of $S=A,H$ couplings vs. $M_S$. (Left) Only $\rho_{i\tau},\rho_{\tau j}\neq0$. (Right) $\rho_{i\tau},\rho_{\tau j}=0$, $\varepsilon\neq0$.} 
    \label{HA_new}
\end{figure}

To compare the constraints from the various observables, we consider two simplifying cases:
\begin{itemize}
    \item In the $\rho$ matrix, only $\rho_{i\tau},\rho_{\tau j}\neq0$. In this case, the two loop contribution vanishes, and $d_\ell\propto {\rm Im}(\rho_{\ell\tau}\rho_{\tau\ell})$, $\Delta a_\mu\propto {\rm Re}(\rho_{\mu\tau}\rho_{\tau\mu})$, and ${\rm BR}(\mu\to e\gamma)\propto(|\rho_{e\tau}\rho_{\tau\mu}|^2+|\rho_{\mu\tau}\rho_{\tau e}|^2)$. The corresponding constraints are presented in Fig.~\ref{HA_new}(Left).
    \item In the $\rho$ matrix, $\rho_{i\tau},\rho_{\tau j}=0$. In this case, the one loop contribution is negligible, and $d_\ell\propto {\rm Re}(\rho_{\ell\ell}\rho_{tt})$, $\Delta a_\mu\propto {\rm Im}(\rho_{\mu\mu}\rho_{tt})$, and ${\rm BR}(\mu\to e\gamma)\propto|\rho_{tt}|^2(|\rho_{ee}|^2+|\rho_{\mu\mu}|^2)$. The corresponding constraints are presented in Fig.~\ref{HA_new}(Right).
\end{itemize}

We learn that for phases in the Yukawa couplings of ${\cal O}(1)$, the projected $d_\mu$ constraint will be competitive with the $\Delta a_\mu$ constraint. The $d_\mu$ constraints will be competitive with the $d_e$ constraints only if the muon-related couplings will be at least four orders of magnitude larger than the corresponding electron-related couplings, well above the MLFV ratio of ${\cal O}(200)$. 

\section{Discussion and Outlook}
\label{sec:con}
The proposal \cite{Adelmann:2025nev} to build an experiment that will be sensitive to an electric dipole moment (EDM) of the muon smaller by three to four orders of magnitude than the current bound \cite{Muong-2:2008ebm}, but still seven order of magnitude larger than the current bound on the EDM of the electron \cite{Roussy:2022cmp}, motivates a study of the potential of such an experiment to find a signal or, alternatively, to put unique constraints on the parameter space of new physics. In this work, we study this question in the framework of new physics that is much heavier than the electroweak scale.

In the context of the Standard Model effective field theory (SMEFT), we identified the dipole operators ${\cal O}_{\mu W}$ and ${\cal O}_{\mu B}$ as the two dimension-six operators where the projected $d_\mu$ sensitivity is the strongest. For four classes of four fermion operators -- ${\cal O}_{luq\mu},\ {\cal O}_{l\mu qu},\ {\cal O}_{l\mu\bar d\bar q}$, and ${\cal O}_{l\mu\bar\mu'\bar l'}$ -- the $d_\mu$ constraints will be stronger than the $d_e$ ones if the Wilson coefficients do not obey the minimal lepton flavor violation (MLFV) relations and the muon-related Wilson coefficients are at least four orders of magnitude larger than the corresponding electron-related ones. For all other dimension-six operators, the projected $d_\mu$ reach is not competitive with the $d_e$ current sensitivity. 

We further examined three representative models of new physics. In a model of vector-like leptons (VLL) that couple dominantly to muons, the constraints from $\Delta a_\mu$, $\Gamma(h\to\mu^+\mu^-)$ and the projected $d_\mu$ are complimentary and competitive. For couplings of ${\cal O}(1)$, the sensitivity of all three measurements is to VLL masses up to ${\cal O}(20\ {\rm TeV})$. In a model of a heavy vector boson $Z^\prime$, $d_\mu$ will explore the regions of parameter space where $Z^\prime$ couples to $\mu\tau$ much more strongly than to $e\tau$, and where the size and the phase of the $Z^\prime\mu\tau$ coupling are ${\cal O}(1)$. For couplings of ${\cal O}(1)$, the sensitivity is to $Z^\prime$ mass up to ${\cal O}(5\ {\rm TeV})$. In a two Higgs doublet model (2HDM), for phases in the Yukawa couplings of the heavy scalars $H$ and $A$ of ${\cal O}(1)$, the projected $d_\mu$ constraint will be competitive with the $\Delta a_\mu$ constraint. The $d_\mu$ constraints will be competitive with the $d_e$ constraints only if the muon-related couplings are at least four orders of magnitude larger than the corresponding electron-related couplings, well above the MLFV ratio of ${\cal O}(200)$. For ${\cal O}(1)$ Yukawa couplings, the sensitivity is to $H,A$ masses up to ${\cal O}(10\ {\rm TeV})$.

We conclude that if the proposed $d_\mu$ experiment will establish an EDM of the muon of order $10^{-22}\ e$ cm, it will imply new physics that couples to leptons at the 10 TeV scale, large CP violation, and significant breaking of minimal flavor violation and other common flavor models. It will also be relevant to the puzzle of the baryon asymmetry of the Universe \cite{Fuchs:2019ore}.  

\section*{Acknowledgements}
YN is supported by a grant from the Minerva Foundation (with funding from the Federal Ministry for Education and Research). 


\begin{thebibliography}{99}

\bibitem{Roussy:2022cmp}
T.~S.~Roussy, L.~Caldwell, T.~Wright, W.~B.~Cairncross, Y.~Shagam, K.~B.~Ng, N.~Schlossberger, S.~Y.~Park, A.~Wang and J.~Ye, \textit{et al.}
``An improved bound on the electron\textquoteright{}s electric dipole moment,''
Science \textbf{381}, no.6653, adg4084 (2023)
[arXiv:2212.11841 [physics.atom-ph]].

\bibitem{Muong-2:2008ebm}
G.~W.~Bennett \textit{et al.} [Muon (g-2)],
``An Improved Limit on the Muon Electric Dipole Moment,''
Phys. Rev. D \textbf{80}, 052008 (2009)
[arXiv:0811.1207 [hep-ex]].


\bibitem{Feng:2001sq}
J.~L.~Feng, K.~T.~Matchev and Y.~Shadmi,
``Theoretical expectations for the muon's electric dipole moment,''
Nucl. Phys. B \textbf{613}, 366-381 (2001)
[arXiv:hep-ph/0107182 [hep-ph]].

\bibitem{Feng:2002wf}
J.~L.~Feng, K.~T.~Matchev and Y.~Shadmi,
``The Measurement of the muon's anomalous magnetic moment isn't,''
Phys. Lett. B \textbf{555}, 89-91 (2003)
[arXiv:hep-ph/0208106 [hep-ph]].

\bibitem{Ema:2021jds}
Y.~Ema, T.~Gao and M.~Pospelov,
``Improved Indirect Limits on Muon Electric Dipole Moment,''
Phys. Rev. Lett. \textbf{128}, no.13, 131803 (2022)
[arXiv:2108.05398 [hep-ph]].

\bibitem{Yamaguchi:2020eub}
Y.~Yamaguchi and N.~Yamanaka,
``Large long-distance contributions to the electric dipole moments of charged leptons in the standard model,''
Phys. Rev. Lett. \textbf{125}, 241802 (2020)
[arXiv:2003.08195 [hep-ph]].

\bibitem{Adelmann:2025nev}
A.~Adelmann, A.~R.~Bainbridge, I.~Bailey, A.~Baldini, S.~Basnet, N.~Berger, L.~Bianco, C.~Calzolaio, L.~Caminada and G.~Cavoto, \textit{et al.}
``A compact frozen-spin trap for the search for the electric dipole moment of the muon,''
Eur. Phys. J. C \textbf{85}, no.6, 622 (2025)
[arXiv:2501.18979 [hep-ex]].

\bibitem{Crivellin:2018qmi}
A.~Crivellin, M.~Hoferichter and P.~Schmidt-Wellenburg,
Phys. Rev. D \textbf{98}, no.11, 113002 (2018)
doi:10.1103/PhysRevD.98.113002
[arXiv:1807.11484 [hep-ph]].

\bibitem{Abada:2024hpb}
A.~Abada and T.~Toma,
``Electric dipole moments of charged leptons in models with pseudo-Dirac sterile fermions,''
JHEP \textbf{08}, 128 (2024)
[arXiv:2405.01648 [hep-ph]].

\bibitem{Hou:2021zqq}
W.~S.~Hou, G.~Kumar and S.~Teunissen,
``Charged lepton EDM with extra Yukawa couplings,''
JHEP \textbf{01}, 092 (2022)
[arXiv:2109.08936 [hep-ph]].

\bibitem{Nakai:2022vgp}
Y.~Nakai, R.~Sato and Y.~Shigekami,
``Muon electric dipole moment as a probe of flavor-diagonal CP violation,''
Phys. Lett. B \textbf{831}, 137194 (2022)
[arXiv:2204.03183 [hep-ph]].

\bibitem{Ellis:2001yza}
J.~R.~Ellis, J.~Hisano, M.~Raidal and Y.~Shimizu,
``Lepton electric dipole moments in nondegenerate supersymmetric seesaw models,''
Phys. Lett. B \textbf{528}, 86-96 (2002)
[arXiv:hep-ph/0111324 [hep-ph]].

\bibitem{Hiller:2010ib}
G.~Hiller, K.~Huitu, T.~Ruppell and J.~Laamanen,
``A Large Muon Electric Dipole Moment from Flavor?,''
Phys. Rev. D \textbf{82}, 093015 (2010)
[arXiv:1008.5091 [hep-ph]].

\bibitem{Hamaguchi:2022byw}
K.~Hamaguchi, N.~Nagata, G.~Osaki and S.~Y.~Tseng,
``Probing new physics in the vector-like lepton model by lepton electric dipole moments,''
JHEP \textbf{01}, 100 (2023)
[arXiv:2211.16800 [hep-ph]].

\bibitem{Dermisek:2023tgq}
R.~Dermisek, K.~Hermanek, N.~McGinnis and S.~Yoon,
``Predictions for muon electric and magnetic dipole moments from $h\to\mu^+\mu^-$ in two-Higgs-doublet models with new leptons,''
Phys. Rev. D \textbf{108}, no.5, 055019 (2023)
[arXiv:2306.13212 [hep-ph]].

\bibitem{Khaw:2022qxh}
K.~S.~Khaw, Y.~Nakai, R.~Sato, Y.~Shigekami and Z.~Zhang,
``A large muon EDM from dark matter,''
JHEP \textbf{02}, 234 (2023)
[arXiv:2212.02891 [hep-ph]].

\bibitem{Altmannshofer:2020ywf}
W.~Altmannshofer, S.~Gori, H.~H.~Patel, S.~Profumo and D.~Tuckler,
``Electric dipole moments in a leptoquark scenario for the $B$-physics anomalies,''
JHEP \textbf{05}, 069 (2020)
[arXiv:2002.01400 [hep-ph]].

\bibitem{Muong-2:2025xyk}
D.~P.~Aguillard \textit{et al.} [Muon g-2],
[arXiv:2506.03069 [hep-ex]].

\bibitem{Aliberti:2025beg}
R.~Aliberti, T.~Aoyama, E.~Balzani, A.~Bashir, G.~Benton, J.~Bijnens, V.~Biloshytskyi, T.~Blum, D.~Boito and M.~Bruno, \textit{et al.}
Phys. Rept. \textbf{1143} (2025), 1-158
doi:10.1016/j.physrep.2025.08.002
[arXiv:2505.21476 [hep-ph]].

\bibitem{MEGII:2025gzr}
K.~Afanaciev \textit{et al.} [MEG II],
[arXiv:2504.15711 [hep-ex]].

\bibitem{Panico:2018hal}
G.~Panico, A.~Pomarol and M.~Riembau,
``EFT approach to the electron Electric Dipole Moment at the two-loop level,''
JHEP \textbf{04}, 090 (2019)
[arXiv:1810.09413 [hep-ph]].

\bibitem{Aebischer:2021uvt}
J.~Aebischer, W.~Dekens, E.~E.~Jenkins, A.~V.~Manohar, D.~Sengupta and P.~Stoffer,
``Effective field theory interpretation of lepton magnetic and electric dipole moments,''
JHEP \textbf{07}, 107 (2021)
[arXiv:2102.08954 [hep-ph]].

\bibitem{Kley:2021yhn}
J.~Kley, T.~Theil, E.~Venturini and A.~Weiler,
``Electric dipole moments at one-loop in the dimension-6 SMEFT,''
Eur. Phys. J. C \textbf{82}, no.10, 926 (2022)
[arXiv:2109.15085 [hep-ph]].

\bibitem{Cirigliano:2005ck}
V.~Cirigliano, B.~Grinstein, G.~Isidori and M.~B.~Wise,
``Minimal flavor violation in the lepton sector,''
Nucl. Phys. B \textbf{728}, 121-134 (2005)
[arXiv:hep-ph/0507001 [hep-ph]].

\bibitem{Cirigliano:2006su}
V.~Cirigliano and B.~Grinstein,
``Phenomenology of minimal lepton flavor violation,''
Nucl. Phys. B \textbf{752}, 18-39 (2006)
[arXiv:hep-ph/0601111 [hep-ph]].

\bibitem{Dery:2013aba}
A.~Dery, A.~Efrati, G.~Hiller, Y.~Hochberg and Y.~Nir,
``Higgs couplings to fermions: 2HDM with MFV,''
JHEP \textbf{08}, 006 (2013)
[arXiv:1304.6727 [hep-ph]].

\bibitem{Fuchs:2019ore}
E.~Fuchs, M.~Losada, Y.~Nir and Y.~Viernik,
``Implications of the Upper Bound on $\boldsymbol{h\to\mu^+\mu^-}$ on the Baryon Asymmetry of the Universe,''
Phys. Rev. Lett. \textbf{124}, no.18, 181801 (2020)
[arXiv:1911.08495 [hep-ph]].

\bibitem{Nir:2024oor}
Y.~Nir and P.~P.~Udhayashankar,
``Lessons from ATLAS and CMS measurements of Higgs boson decays to second generation fermions,''
JHEP \textbf{06}, 049 (2024)
[arXiv:2404.16545 [hep-ph]].

\bibitem{Fuchs:2020uoc}
E.~Fuchs, M.~Losada, Y.~Nir and Y.~Viernik,
``$CP$ violation from $\tau$, $t$ and $b$ dimension-6 Yukawa couplings - interplay of baryogenesis, EDM and Higgs physics,''
JHEP \textbf{05}, 056 (2020)
[arXiv:2003.00099 [hep-ph]].

\bibitem{CMS:2020xwi}
A.~M.~Sirunyan \textit{et al.} [CMS],
``Evidence for Higgs boson decay to a pair of muons,''
JHEP \textbf{01}, 148 (2021)
[arXiv:2009.04363 [hep-ex]].

\bibitem{ATLAS:2020fzp}
G.~Aad \textit{et al.} [ATLAS],
``A search for the dimuon decay of the Standard Model Higgs boson with the ATLAS detector,''
Phys. Lett. B \textbf{812}, 135980 (2021)
[arXiv:2007.07830 [hep-ex]].

\bibitem{Vives:2025clr}
O.~Vives and N.~Valori,
[arXiv:2505.06345 [hep-ph]].

\bibitem{Altmannshofer:2020shb}
W.~Altmannshofer, S.~Gori, N.~Hamer and H.~H.~Patel,
Phys. Rev. D \textbf{102}, no.11, 115042 (2020)
doi:10.1103/PhysRevD.102.115042
[arXiv:2009.01258 [hep-ph]].

\bibitem{Altmannshofer:2025nsl}
W.~Altmannshofer, B.~Assi, J.~Brod, N.~Hamer, J.~Julio, P.~Uttayarat and D.~Volkov,
JHEP \textbf{06}, 156 (2025)
doi:10.1007/JHEP06(2025)156
[arXiv:2410.17313 [hep-ph]].

\end{thebibliography}

\end{document}